\documentclass[twocolumn]{aastex62}

\newcommand{\hi}{H\,{\sc i}}

\graphicspath{{./}{figures/}}
\usepackage{float}
\usepackage[caption = false]{subfig}
\usepackage{gensymb}
\usepackage{hyperref} 
\usepackage{lineno}
\usepackage{scalerel}

\shorttitle{DELVE-DEEP: NGC 55 Survey}
\begin{document}

%\title{Searching for Dwarf Galaxy Satellites of NGC 55 with the DELVE-DEEP Survey}
\title{DELVE-DEEP Survey: The Faint Satellite System of NGC~55}
%\author[]{}
%\affil{}
%\email{}

\correspondingauthor{Jonah Medoff}
\email{jmedoff@fas.harvard.edu}

\author[0000-0003-1162-7346]{Jonah~Medoff}
\affiliation{Department of Astronomy and Astrophysics, University of Chicago, Chicago, IL 60637, USA}
\affiliation{Department of Physics, Harvard University, 17 Oxford St, Cambridge, MA 02138, USA}

\author[0000-0001-9649-4815]{\textsc{Bur\c{c}{\rlap{\.}\i}n~Mutlu-Pakd{\rlap{\.}\i}l}}
\affil{Department of Physics and Astronomy, Dartmouth College, Hanover, NH 03755, USA}

\author[0000-0002-3936-9628]{Jeffrey~L.~Carlin}
\affiliation{Rubin Observatory/AURA, 950 North Cherry Avenue, Tucson, AZ, 85719, USA}

\author[0000-0001-8251-933X]{Alex~Drlica-Wagner}
\affiliation{Department of Astronomy and Astrophysics, University of Chicago, Chicago, IL 60637, USA}
\affiliation{Fermi National Accelerator Laboratory, P. O. Box 500, Batavia, IL 60510, USA}
\affiliation{Kavli Institute for Cosmological Physics, University of Chicago, Chicago, IL 60637, USA}

\author[0000-0002-9599-310X]{Erik~J.~Tollerud}
\affiliation{Space Telescope Science Institute, 3700 San Martin Drive, Baltimore, MD 21218, USA}

\author[0000-0001-9775-9029]{Amandine~Doliva-Dolinsky}
\affiliation{Department of Physics and Astronomy, University of Tampa, 401 West Kennedy Boulevard, Tampa, FL 33606, USA}
\affiliation{Department of Physics and Astronomy, Dartmouth College, Hanover, NH 03755, USA}

\author[0000-0003-4102-380X]{David~J.~Sand}
\affiliation{Steward Observatory, University of Arizona, 933 North Cherry Avenue, Tucson, AZ 85721-0065, USA}

\author[0000-0002-9144-7726]{Clara~E.~Mart\'inez-V\'azquez}
\affiliation{International Gemini Observatory/NSF NOIRLab, 670 N. A’ohoku Place, Hilo, Hawai’i, 96720, USA}

\author[0000-0003-1479-3059]{Guy~S.~Stringfellow}
\affiliation{Center for Astrophysics and Space Astronomy, University of Colorado, 389 UCB, Boulder, CO 80309-0389, USA}

\author[0000-0003-1697-7062]{William~Cerny}
\affiliation{Department of Astronomy, Yale University, New Haven, CT 06520, USA}

\author[0000-0002-1763-4128]{Denija~Crnojevi\'c}
\affiliation{Department of Physics and Astronomy, University of Tampa, 401 West Kennedy Boulevard, Tampa, FL 33606, USA}

\author[0000-0001-6957-1627]{Peter S. Ferguson}
\affiliation{DIRAC Institute, Department of Astronomy, University of Washington, 3910 15th Ave NE, Seattle, WA, 98195, USA}

\author[0000-0001-8245-779X]{Catherine E. Fielder}
\affiliation{Steward Observatory, University of Arizona, 933 North Cherry Avenue, Tucson, AZ 85721-0065, USA}

\author[0000-0001-5143-1255]{Astha Chaturvedi}
\affiliation{Department of Physics, University of Surrey, Guildford GU2 7XH, UK}

\author[0000-0002-3204-1742]{Nitya~Kallivayalil}
\affiliation{Department of Astronomy, University of Virginia, 530 McCormick Road, Charlottesville, VA 22904, USA}

\author[0000-0002-8282-469X]{Noelia~E.~D.~No\"el}
\affiliation{Department of Physics, University of Surrey, Guildford GU2 7XH, UK}

\author[0000-0003-4341-6172]{Kathy~Vivas}
\affiliation{Cerro Tololo Inter-American Observatory/NSF NOIRLab, Casilla 603, La Serena, Chile}

\author[0000-0002-7123-8943]{Alistair~R.~Walker}
\affiliation{Cerro Tololo Inter-American Observatory/NSF NOIRLab, Casilla 603, La Serena, Chile}

\author[0000-0002-6904-359X]{Monika~Adam\'ow}
\affiliation{Center for Astrophysical Surveys, National Center for Supercomputing Applications, 1205 West Clark St., Urbana, IL 61801, USA}

\author[0000-0003-4383-2969]{Clecio~R.~Bom}
\affiliation{Centro Brasileiro de Pesquisas F\'isicas, Rua Dr. Xavier Sigaud 150, 22290-180 Rio de Janeiro, RJ, Brazil}

\author[0000-0002-3690-105X]{Julio~A.~Carballo-Bello}
\affiliation{Instituto de Alta Investigaci\'on, Universidad de Tarapac\'a, Casilla 7D, Arica, Chile}

\author[0000-0003-1680-1884]{Yumi~Choi}
\affiliation{NSF NOIRLab, 950 N. Cherry Ave., Tucson, AZ 85719, USA}

\author[0000-0003-0105-9576]{Gustavo~E.~Medina}
\affiliation{Department of Astronomy and Astrophysics, University of Toronto, 50 St. George Street, Toronto ON, M5S 3H4, Canada}

\author[0000-0001-9438-5228]{Mahdieh~Navabi}
\affiliation{Department of Physics, University of Surrey, Guildford GU2 7XH, UK}

\author[0000-0002-6021-8760]{Andrew~B.~Pace}
\affiliation{Department of Astronomy, University of Virginia, 530 McCormick Road, Charlottesville, VA 22904, USA}

\author[0000-0001-5805-5766]{Alex~H.~Riley}
\affiliation{Institute for Computational Cosmology, Department of Physics, Durham University, South Road, Durham DH1 3LE, UK}

\author[0000-0002-1594-1466]{Joanna~D.~Sakowska}
\affiliation{Department of Physics, University of Surrey, Guildford GU2 7XH, UK}
\affiliation{Instituto de Astrofísica de Andalucía, CSIC, Glorieta de la Astronom\'\i a,  E-18080 Granada, Spain}

\collaboration{(DELVE Collaboration)}
%\affiliation{}

%(DELVE Collaboration)

\begin{abstract}

We report the first comprehensive census of the satellite dwarf galaxies around NGC~55 ($2.1$~Mpc) as a part of the DECam Local Volume Exploration DEEP (DELVE-DEEP) survey. NGC~55 is one of four isolated, Magellanic analogs in the Local Volume around which DELVE-DEEP aims to identify faint dwarfs and other substructures. We employ two complementary detection methods: one targets fully resolved dwarf galaxies by identifying them as stellar overdensities, while the other focuses on semiresolved dwarf galaxies, detecting them through shredded unresolved light components. As shown through extensive tests with injected galaxies, our search is sensitive to candidates down to $M_V \lesssim -6.6$ and surface brightness $\mu \lesssim 28.5$ mag arcsec$^{2}$, and $\sim 80\%$ complete down to $M_V \lesssim -7.8$. We do not report any new confirmed satellites beyond two previously known systems, ESO 294-010 and NGC~55-dw1. We construct the satellite luminosity function of NGC~55 and find it to be consistent with the predictions from cosmological simulations. As one of the first complete luminosity functions for a Magellanic analog, our results provide a glimpse of the constraints on low-mass-host satellite populations that will be further explored by upcoming surveys, such as the Vera C. Rubin Observatory's Legacy Survey of Space and Time.

\end{abstract}

\section{Introduction}
\label{sec:intro}

The $\Lambda$Cold Dark Matter ($\Lambda$CDM) cosmological model has been well assessed on large scales, with observations of the large-scale structure of the Universe and cosmic microwave background providing strong support for its predictions (e.g., \citealt{DES_2018,Planck_2020}). In comparison, observational constraints on the small-scale distribution of dark matter are relatively limited \citep{Bullock17}. 
Dwarf galaxies, being the most ancient, metal-poor, %chemically pristine, 
and dark-matter-dominated galaxies in the Universe, serve as excellent probes of early galaxy formation and small-scale dark matter distribution \citep{Simon2019}, thus providing a means of exploring $\Lambda$CDM on small scales \citep{Bullock17}. Within the last 20 years, dozens of dwarf galaxies have been discovered around the Milky Way (MW; e.g, \citealt{Willman_2005}; \citealt{Belokurov_2006}; \citealt{Simon_2007}; \citealt{Bechtol_2015}; \citealt{Drlica_wagner_2015}; \citealt{Koposov_2015}; \citealt{Mau20}; \citealt{Cerny_2021a, Cerny_2021b, Cerny_2023}; \citealt{Homma_2024}) and other MW-mass hosts (e.g., M31: \citealt{Martin2013}, \citealt{Doliva-Dolinsky_2022, Doliva-Dolinsky_2023}, \citealt{Arias_2025}, \citealt{Smith_2025}; M81: \citealt{Chiboucas2009}; Cen~A: \citealt{Crnojevic16, Crnojevic19}; M94: \citealt{Smercina18}; M101: \citealt{Bennet19, Bennet20}; NGC~253: \citealt{Mutlu-Pakdil_2024}; ELVES: \citealt{Carlsten_2022}; SAGA: \citealt{Mao_2024}).

\begin{table*}[t]
\label{table:table1}
\centering
\caption{Properties of NGC 55, ESO 294-010, and NGC~55-dw1}
\begin{tabular}{c c c c c} 
 \hline\hline
 Parameter & NGC 55 & ESO 294-010 & NGC~55-dw1 & References \\ [0.5ex] 
 \hline
 R.A. (deg) &  00$^{\text{h}}$14$^{\text{m}}$53$^{\text{s}}$\rlap{\hspace{-0.3em}.}6 & 00$^{\text{h}}$26$^{\text{m}}$33$^{\text{s}}$\rlap{\hspace{-0.3em}.}5 & 00$^{\text{h}}$15$^{\text{m}}$28$^{\text{s}}$\rlap{\hspace{-0.3em}.}8 & NED (1) \\ 
 Decl. (deg) & -39\degree11'47''\rlap{\hspace{-0.4em}.}9 & -41\degree51'18''\rlap{\hspace{-0.4em}.}2 & -38\degree25'08''\rlap{\hspace{-0.4em}.}4 & NED (1) \\
 Distance (Mpc) & 2.1 & 2.0 & 2.2 & (2, 3, 1) \\
 Radial Velocity (km/s) & 129 & 117 & ... & (3)\\
 Stellar Mass ($M_\odot$) & $3.0\times10^{9}$ & $2.7\times10^{6}$ & ... & (4, 3) \\
 Virial Radius (kpc) & 120 & ... & ... & (5) \\
 $M_V$ (mag) & -18.6 & -11.3 & -8.0 & (6, 7, 1) \\
 $r_h$ (pc) & 3916 & 248 & 2200 & (8, 3, 1) \\
 $r_h$ (arcsec) & 385 & 24 & 220 & (8, 3, 1) \\
 $\mu_{\text{eff}}$ (mag arcsec$^{-2}$) & 22.8 & 24.1 & 32.3 & (8, 7, 3, 1) \\
 $D_{\text{proj}}$ (kpc) & N/A & 120.8 & 30.2 & NED (1) \\[1ex]
 %Halo Mass ($M_\odot$) & $2\times10^{10}$ & ... & ... & (6) \\ [1ex] 
 \hline
 %\multicolumn{5}{c}{\hspace{1pt}}\\
 \multicolumn{5}{c}{} \\[-7pt]
 %\multicolumn{5}{c}{\parbox{0.75\textwidth}{\centering * $r_h$ (arcsec) for NGC~55 represents the full radius in arcsec as listed by NED, as no half-light radius was found. The corresponding $r_h$ (pc) and $\mu_{eff}$ were calculated using this $r_h$~(arcsec) and distance to NGC~55 (2.1 Mpc), and thus represent the full radius in pc and average surface brightness, respectively.}}\\
 %\multicolumn{5}{c}{\hspace{1pt}}\\
 \multicolumn{5}{c}{} \\[-7pt]
 \multicolumn{5}{c}{\parbox{0.75\textwidth}{\centering \textbf{References:} (1) \citealt{McNanna2024}, (2) \citealt{Dalcanton_2009}, (3) \citealt{McConnachie2012}, (4) \citealt{Dooley2017b}, (5) \citealt{Mutlupakdil21}, (6) \citealt{GALEX_2007}, (7) \citealt{Karachentsev_2002}, (8) \citealt{Moustakas_2023}
 %(6) \citealt{Westmeier_2013}}
 }}
\end{tabular}
\end{table*}

In order to thoroughly assess the small-scale predictions of $\Lambda$CDM, we need to explore satellite systems in host environments with lower masses than the MW to better understand how astrophysical processes, such as the effects of reionization, tidal and ram pressure stripping, and host infall time, affect satellite populations (\citealt{Drlica-Wagner2021}). Ideally, we could study these effects using the Magellanic Clouds (MCs), but, due to their location within the MW halo, the association of some satellites between the MCs and MW remains ambiguous (e.g., \citealt{Jethwa_2016}; \citealt{Patel_2020}; \citealt{Cerny_2021a}). Furthermore, satellites that are known to be bound to the MCs have likely experienced recent interactions with the MW, complicating the interpretation of their gas contents, stellar populations, and orbital dynamics. As a result, we must search for satellites of isolated, MC-mass hosts beyond the Local Group. Several studies in recent years have already begun to perform such searches (e.g., \citealt{Sand15b, sand_2024}; MADCASH: \citealt{Carlin16, carlin21, Carlin_2024}; LBT-SONG: \citealt{Davis21}; ELVES-Dwarf: \citealt{Li_2025}; ID-MAGE: \citealt{Congreve_2025}), discovering several dwarf satellites around MC analogs in the Local Volume, such as NGC~3109, NGC~4124, and NGC~628, and the first complete satellite census of an MC-mass host (NGC~2403) was recently presented in \citet{Carlin_2024}. To sufficiently test $\Lambda$CDM on small scales, however, it is necessary to study the satellite populations of a larger sample of MC-mass hosts.

As part of the DEEP component of the DECam Local Volume Exploration (DELVE-DEEP; \citealt{Drlica-Wagner2021, Drlica-Wagner_2022}) survey, we perform a comprehensive dwarf satellite search around NGC~55, a barred spiral galaxy located at $\sim$2.1 Mpc, with a mass and morphology similar to the LMC (\citealt{de_Vaucoleurs_1991}; \citealt{Dalcanton_2009}; \citealt{Westmeier_2013}). NGC~55 is considered to be a member of the Sculptor Group, but given the weak gravitational bound of the Group \citep{Jerjen1998,Karachentsev_2003} and the low tidal index\footnote{The tidal index, $\Theta_5$, is a measure of the cumulative gravitational influence of a galaxy's five most significant neighbors (see \citealt{Karachentsev2013} for definition).} of NGC 55 ($\Theta_5 = 0.1$; \citealt{Karachentsev2013}), it is likely to only be loosely bound to this group. Within its local environment, NGC~55 has several nearby neighbors: NGC~300 ($D_{\text{proj}} \sim 278.2$~kpc) and two dwarf spheroidal (dSph) galaxies, ESO~410-005 ($D_{\text{proj}} \sim 244.8$~kpc) and ESO~294-010 ($D_{\text{proj}} \sim 120.8$~kpc). The latter of these dSphs is located at the edge of the virial radius of NGC~55 (see Table \ref{table:table1}) and is a likely satellite companion, judging by its tip of the red-giant-branch (TRGB) distance and radial velocity ($D = 2.0$ Mpc; $v_r = 117$ km/s; \citealt{McConnachie2012}).

Using the high-resolution, dark-matter-only \textit{Caterpillar} suite of simulations \citep{Griffen_2016} and stellar mass-halo mass relation of \citet{Garrison-Kimmel_2017}, \citet{Dooley2017b} predicts that an LMC-mass host like NGC~55 should host a total of 2-6 satellite galaxies with $M_V \le -7$. In addition to ESO 294-010 ($M_V \sim -11.3$; \citealt{Karachentsev_2002}), a second, more diffuse dwarf satellite of NGC 55, NGC~55-dw1 ($M_V \sim -8.0$), was recently identified using Dark Energy Survey (DES) Year 6 and DELVE-DEEP data and presented in \citet{McNanna2024}. Taking these known dwarfs into account, the predictions of \citet{Dooley2017b} suggest that there should be 0-4 remaining dwarfs detectable within our data. The basic properties of NGC~55 and its two known satellites are listed in Table~\ref{table:table1}.

Previous satellite searches around isolated, low-mass hosts have utilized either resolved star detection methods (e.g., \citealt{Carlin16, carlin21}) or integrated light detection methods (e.g., \citealt{Davis21}). While these two methods have been effective, neither method, on their own, spans the entire size-luminosity parameter space of known dwarf galaxies. To address this, we implement a new approach that combines both a resolved and ``semiresolved'' search for dwarf galaxies (see Section~\ref{subsec:semi-resolved search} for a definition of semiresolved). Through extensive tests with injected galaxies (see Section~\ref{sec:simulations}), we show that this method is sensitive to a parameter space of dwarfs that extends well beyond those of previous satellite searches (at the distance of NGC~55).

In this work, we present the results of our systematic NGC~55 satellite search, along with our resulting satellite luminosity function (LF) for NGC~55 down to a limit of $M_V\sim-7$. Section~\ref{sec:delve_deep} presents an overview of the DELVE-DEEP survey. Section~\ref{sec: dwarf_search} outlines our approach for detecting dwarf galaxy satellites and the results obtained. Section~\ref{sec:simulations} details our methodology for performing artificial dwarf galaxy simulations, along with our assessment of the detection sensitivity and completeness. In Section~\ref{sec:discussion}, we present the satellite LF of NGC~55 and place its satellite system in context with other known systems, and, finally, in Section~\ref{sec:conclusions}, we summarize our conclusions.

\section{The DELVE-DEEP Survey and Data Processing} \label{sec:delve_deep}

DELVE is a multicomponent survey using the Dark Energy Camera \citep{Flaugher_2015A} on the Blanco 4m Telescope at the Cerro Tololo Inter-American Observatory that aims to characterize satellite galaxy populations across a range of environments in the Local Volume \citep{Drlica-Wagner2021}. DELVE-DEEP is one of the three components of this survey that specifically targets isolated MC analogs beyond the Local Group to study their dwarf satellite populations and other substructures within their halo. DELVE-DEEP obtained 135~deg$^2$ of deep imaging in the $g$ and $i$ optical bands around four MC-mass host galaxies at 1.4-2.1~Mpc: two LMC-mass analogs (NGC~55 and NGC~300) and two SMC-mass analogs (Sextans~B and IC~5152). This program is complementary to the MADCASH survey \citep{Carlin16, carlin21, Carlin_2024}, which aims to investigate the satellite populations and substructures of seven additional MC-mass hosts in the Local Volume (two of which have been completed; see \citealt{Carlin_2024} and \citealt{Doliva-Dolinsky_2025}). Recently, several stellar substructures were discovered around NGC~300 using DELVE-DEEP data and are presented in \citet{Fielder_2025}. NGC~55, however, is the first of the four DELVE-DEEP targets around which we have performed a comprehensive dwarf galaxy satellite search. 

\begin{figure*}[t]
    \centering
    \includegraphics[width = \textwidth]{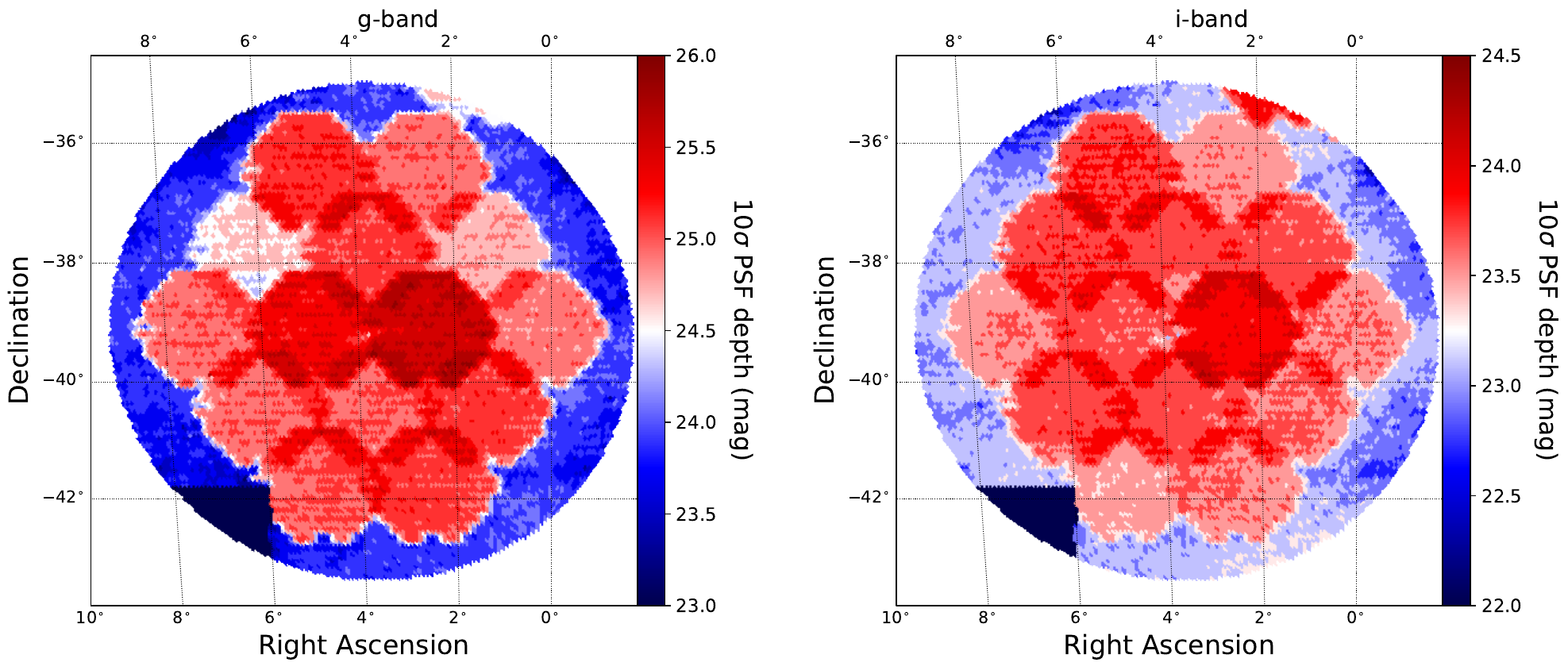}
    \caption{10$\sigma$ PSF depth of DECam photometry around NGC~55 in $g$ and $i$ optical bands. 10$\sigma$ PSF depth is defined as the PSF magnitude at which the signal-to-noise ratio is equal to 10. The central red regions correspond to fields with both DELVE-DEEP and DES imaging, while the outer blue regions correspond to fields with only DES imaging. The field in the bottom left corner was imaged as part of the DES supernova program and was excluded from our dataset. The total region shown spans a projected physical radius of $\sim$150~kpc.}
    \label{fig:ngc55_depth}
\end{figure*}

The DELVE-DEEP observations of NGC~55 were collected on 17 distinct nights between 2019 August and 2021 July and performed over 14 distinct fields within a 3.25~deg radius from the galactic center, which roughly corresponds to the virial radius of NGC~55 ($r_{\text{vir}} \sim 120$~kpc). Within each field, 12 $g$-band exposures and 7 $i$-band exposures were taken, each with an exposure time of 300~s and good seeing (FWHM $< 1.0$ arcsec). Combined with the DES exposures in these fields, this gives a total of 15$\times$300~s exposures in $g$ and 10$\times$300~s exposures in $i$ (\citealt{Drlica-Wagner2021}). 
%Additional exposures were taken when seeing conditions were poor, ensuring a total of at least 12 $g$-band exposures and 7 $i$-band exposures with good transparency in each field. 
Across all fields, the median point-spread function (PSF) FWHM was 0.97\arcsec~in $g$ and 0.82\arcsec~in $i$. Additionally, each observation within each field was dithered by 4\arcmin~in R.A. and 2\arcmin~in decl. to mitigate chip gaps in the images and improve the uniformity of depth. 

Image processing was performed via the DES Data Management pipeline (DESDM; \citealt{Morganson_2018}; for a description of how the DESDM pipeline is applied to DELVE data, see \citealt{tan2024}), which 
%catalogs images using
is based on \texttt{SourceExtractor} \citep{Bertin1996}. Images of the sky are broken up into 0.73~deg$^2$ regions called ``tiles'' \citep{Morganson_2018}. Despite DELVE-DEEP only using the $g$ and $i$ bands, our data processing requires both the detection image (which is a composite of $g$-, $r$-, and $i$-band tiles) and the single-band image ($g$ or $i$) for each field. \texttt{SourceExtractor} uses the detection image to identify sources and then performs forced photometry using the single-band image, which enhances its overall detection sensitivity. The same procedure is implemented in our dwarf galaxy simulations (see Section~\ref{sec:simulations}) to ensure consistency in detecting both real and artificial sources.

In addition to DESDM processing, extinction corrections were applied to all $g$- and $i$-band magnitudes according to the procedure described in \citet{Drlica-Wagner2021, Drlica-Wagner_2022}. Contrary to previous dwarf galaxy searches (e.g., \citealt{Carlin16}), star-galaxy separation was not applied to the data. Our artificial dwarf galaxy simulations (see Section~\ref{sec:simulations}) reveal that, at the distance of NGC~55, blending and crowding of stars in the central regions of dwarf galaxies, combined with underlying diffuse light, lead to uncertain photometry that makes star-galaxy separation unreliable. As a result, many of these stars are classified as galaxies, which would cause many dwarf galaxies to go undetected in our search if star-galaxy separation was applied. Thus, to optimize the completeness of our dwarf search, we find it more effective not to apply star-galaxy separation. While this choice increases the number of false detections in our dwarf galaxy search, we mitigate this by visually inspecting each candidate (see Section~\ref{subsec:visual_inspection}). Although other PSF-fitting photometry programs, such as DAOPHOT \citep{Stetson87}, could potentially provide more reliable star-galaxy separation, applying these methods to the entire survey would have been significantly more time-consuming.

DELVE-DEEP has achieved average 10$\sigma$ depths of $g = 25.1$~mag and $i = 23.7$~mag around NGC~55 (where 10$\sigma$ depth is defined as the PSF magnitude at which the signal-to-noise ratio is equal to 10; see Figure~\ref{fig:ngc55_depth}), which is $\sim 1$ mag deeper than the corresponding DES imaging in this region (\citealt{McNanna2024}). Additionally, using artificial star tests (discussed in more detail in Appendix~\ref{appendix b}), we determine that our observations are 90\% complete to depths of $g_{90\%} = 25.3$~mag and $i_{90\%} = 24.6$~mag, providing sensitivity to resolved stars roughly 1.5 mag fainter than the TRGB ($m_g^{\mathrm{TRGB}} = 24.3$;  $m_i^{\mathrm{TRGB}} = 22.6$).

\section{Dwarf Satellite Search} \label{sec: dwarf_search}

In our dwarf search around NGC~55, we take a new approach and implement a pipeline that combines two distinct dwarf galaxy detection methods: one that detects dwarf galaxies containing primarily resolved red-giant-branch (RGB) sources (Section~\ref{subsec:resolved search}), and one that detects dwarf galaxies containing a mix of resolved sources and shredded, unresolved light components (we refer to these dwarf galaxies as ``semiresolved''; Section~\ref{subsec:semi-resolved search}). 

\subsection{Resolved Search}
\label{subsec:resolved search}

\begin{figure}[t]
    \centering
    \includegraphics[width = \columnwidth]{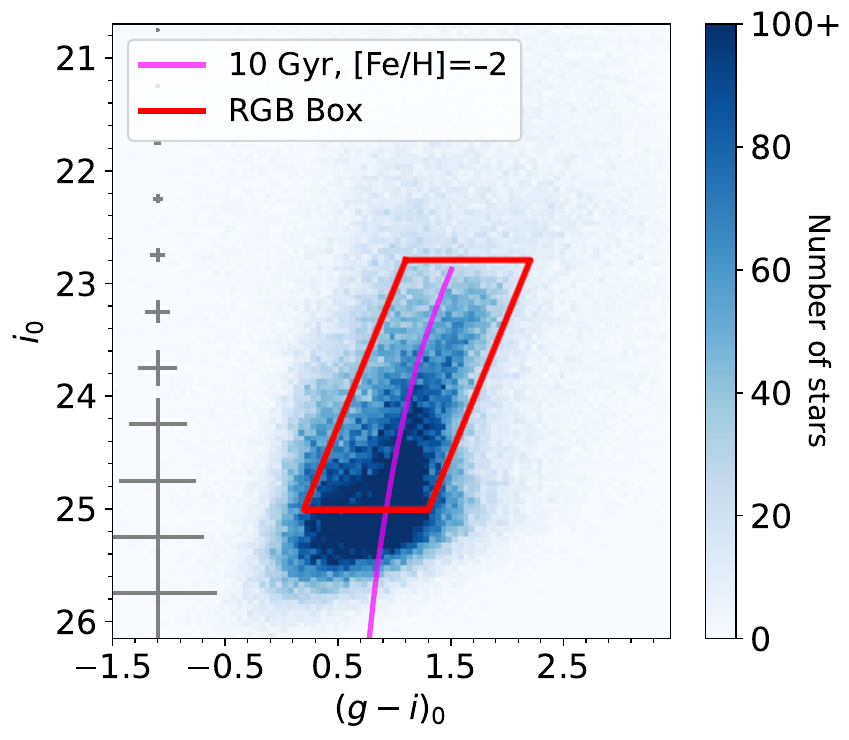}
    \caption{Binned color-magnitude diagram of the central region of NGC~55 with the RGB selection box used in our dwarf galaxy search displayed in red. This RGB box covers $i$-band magnitudes of $22.8 \leq i \leq 25.0$. An isochrone corresponding to stellar populations with age $= 10$ Gyr and metallicity [Fe/H] $=-2$ at the distance of NGC~55 is shown in magenta, which was generated using the Dartmouth Stellar Evolution Database \citep{Dotter2008}. Magnitude uncertainties (which come from the \texttt{SourceExtractor} parameter, \texttt{MAG\_ERR}) are given in gray on the left.}
    \label{fig:cmd}
\end{figure}

Following previous dwarf galaxy searches (e.g., \citealt{Martin2013}; \citealt{Crnojevic16}; \citealt{McQuinn_2023, McQuinn_2024}; \citealt{Carlin_2024}; \citealt{McNanna2024}; \citealt{Mutlu-Pakdil_2024}), one way we identify dwarf galaxies is by using resolved RGB stars. This is because dwarf galaxies primarily consist of old, metal-poor stars, which have evolved off the main sequence onto the RGB. We identify potential RGB stars in our data based on the color-magnitude selection box shown in Figure~\ref{fig:cmd}. The boundaries of this selection box were chosen to cover all of the sources in our data along the RGB and are consistent with an isochrone of stellar populations with age $= 10$ Gyr and [Fe/H] $=-2$ \citep{Dotter2008}.

\begin{figure*}
    \centering
    \includegraphics[width = \textwidth]{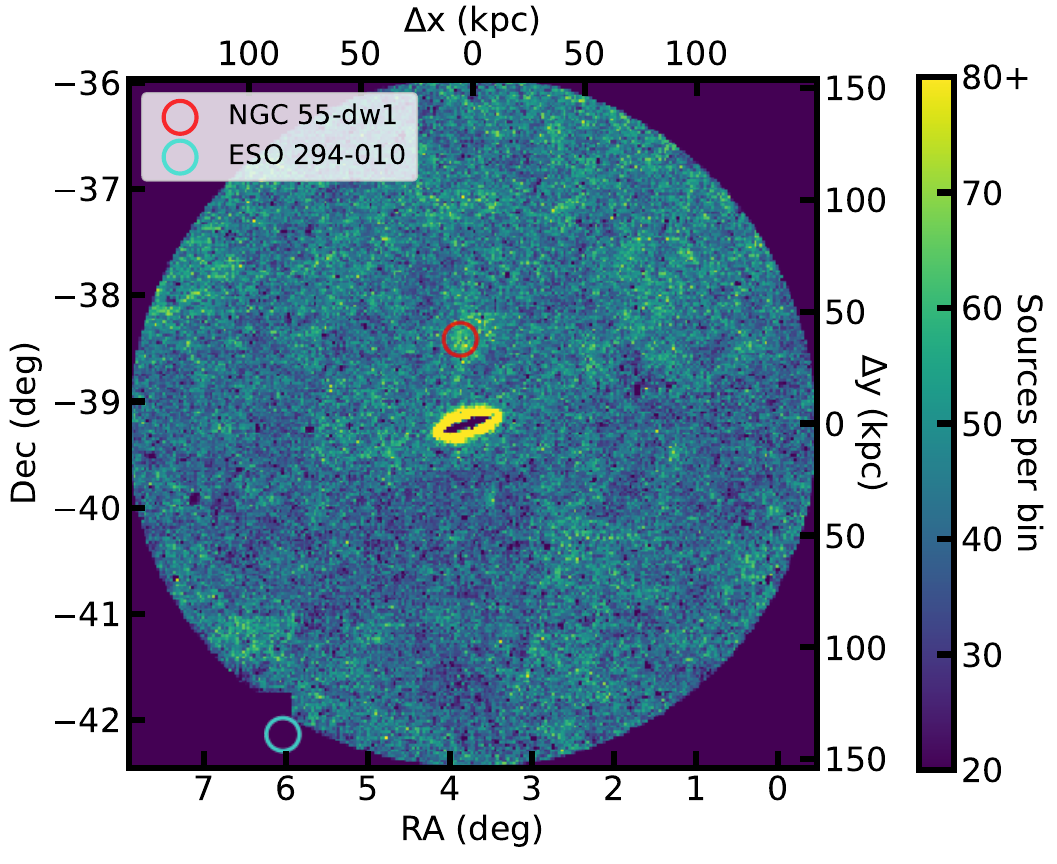}
    \caption{RGB density map of DELVE-DEEP's NGC~55 footprint (rad = 3.25 deg) with binsize = 1.5~arcmin$^2$. The color bar indicates the number of sources per bin. The two previously known NGC~55 satellites, NGC~55-dw1 and ESO~294-010, plotted in red and cyan, respectively. We note that, while ESO~294-010 is located slightly outside the DELVE-DEEP footprint, it lies just at the edge of the virial radius of NGC~55.}
    \label{fig:density_map}
\end{figure*}

The sources within the RGB selection box are divided into 1.5 arcmin$^2$ spatial bins, and for each bin, mean and standard deviation background bin values (\texttt{MEAN\_BKG\_BIN} and \texttt{STD\_BKG\_BIN}) are calculated using the surrounding annulus of bins, with inner radius $2*\texttt{binsize}$ and outer radius $3*\texttt{binsize}$. The resulting binned density map (see Figure~\ref{fig:density_map}) is run through a \texttt{photutils} function called \texttt{find$\_$peaks} (\citealt{astropy13}), with $\texttt{threshold} = \texttt{MEAN\_BKG\_BIN} + 3*\texttt{STD\_BKG\_BIN}$ (i.e., 3$\sigma$ above the mean background value) and $\texttt{box\_size} = 3$, which roughly corresponds to the expected size, in bin length, of a dwarf satellite of NGC~55. This function outputs 419 overdensities, or ``peaks'' (which include the previously discovered satellite, NGC~55-dw1), although many of these are false detections (e.g., overdensities of sources due to a background galaxy, a bright foreground star, or a background galaxy cluster), which are later filtered out through systematic visual inspection (see Section~\ref{subsec:visual_inspection}).

\subsection{Semiresolved Search} \label{subsec:semi-resolved search}

In addition to fully resolved dwarf galaxies, we expect some dwarfs at the distance of NGC~55 to potentially contain a combination of resolved and unresolved stars. Consequently, such systems could go undetected by our resolved search method. These dwarf galaxies (which we refer to as ``semiresolved'' dwarfs) contain central regions of blended, unresolved light surrounded by more dispersed, resolved stars in the outskirts.

Through our artificial dwarf galaxy simulations (see Section~\ref{sec:simulations}), we find that \texttt{SourceExtractor} ``shreds'' the central, unresolved regions of these semiresolved dwarfs into several, distinct sources, each of which exhibits low surface brightness (LSB) and a large flux radius (for a definition and discussion of ``shredding'' in unresolved sources, see \citealt{Prole_2018}). Since these sources cannot be detected by our resolved search, we implement a second detection method that targets overdensities of LSB sources. Note that this methodology differs from previous unresolved dwarf galaxy searches (e.g., \citealt{Davis21}), which target single LSB sources, as opposed to shredded, LSB overdensities. 

To select LSB sources, we utilize LSB selection criteria adapted from those listed in \citet{Tanoglidis2021}. These criteria rely mostly on the \texttt{SourceExtractor} parameters, \texttt{FLUX\_RADIUS} (defined as the radius of a circular isophote containing half of a source's flux) and \texttt{MU\_EFF\_MODEL} (effective surface brightness), and our corresponding selection cuts for these two parameters are:

\noindent \texttt{\small (FLUX\_RADIUS\_G > 5) \& (FLUX\_RADIUS\_G < 20)},\\
\noindent \texttt{\small (FLUX\_RADIUS\_I > 5) \& (FLUX\_RADIUS\_I < 20)},\\
\noindent \texttt{\small (MU\_EFF\_MODEL\_G > 24.2) \& (MU\_EFF\_MODEL\_G < 31.2).}

We test several different selection cuts for these parameters, but find that these limits maximize the number of detected artificial dwarfs (see Section~\ref{sec:simulations}) without resulting in too many false detections in our real dwarf galaxy search. Our other LSB selection criteria come directly from \citet{Tanoglidis2021}, but, in comparison, have relatively little effect on the number of sources detected. These criteria are as follows:

\noindent \texttt{\small (EXTENDED\_CLASS\_G != 0) \& (EXTENDED\_CLASS\_I != 0)},\\
\noindent \texttt{\small (SPREAD\_MODEL\_I + 5/3*SPREADERR\_MODEL\_I > 0.007)},\\
\noindent \texttt{\small (MAG\_AUTO\_G - MAG\_AUTO\_I > -0.1)},\\
\noindent \texttt{\small (MAG\_AUTO\_G - MAG\_AUTO\_I < 1.4)}.\footnote{See \citet{Drlica-Wagner_2022} Eq.~2 and \citet{Desai_2012} for definitions of \texttt{EXTENDED\_CLASS} and \texttt{SPREAD\_MODEL}, respectively. \texttt{MAG\_AUTO} is the automatic aperture magnitude calculated by \texttt{SourceExtractor}.}

In addition to these LSB selection cuts, we also apply color and magnitude cuts similar to the RGB selection in Section \ref{subsec:resolved search}, though these cuts have a negligible effect on the results of our dwarf search.

After applying these cuts, we bin the data into 1.5~arcmin$^2$ bins and apply the \texttt{find$\_$peaks} function with the same \texttt{threshold} and \texttt{box$\_$size} parameters as used in the analysis described in Section~\ref{subsec:resolved search}. This resulted in a list of 1,535 overdensities, which is significantly more than the 419 overdensities detected by the resolved method, since the LSB search picks up more false detections such as the halos of bright foreground stars and background galaxies. These detections are then filtered out through visual inspection.

It is important to note that, since star-galaxy separation is not applied, the resolved search includes all sources that are present in the semiresolved search. One might therefore expect that any overdensity detected in the semiresolved search would also appear in the resolved search. This is not the case, however, since the resolved search has a much higher average number of sources per bin, which reduces the relative significance of semiresolved overdensities, making them less likely to be detected.

\subsection{Visual Inspection}
\label{subsec:visual_inspection}

Our dwarf search yields 419 candidates through the resolved detection method and 1,535 candidates through the semiresolved detection method, which, after removing duplicates, results in a total of 1,943 potential dwarf galaxy candidates. Visual inspection of these candidates was carried out using the citizen science platform, Zooniverse, with three members of our team classifying each candidate as either a potential dwarf galaxy worthy of further follow-up observation or as a false detection. For each candidate, we inspected a color-magnitude diagram, spatial distribution, radial density profile, LF, and color image (similar to the diagnostic plots described in \citealt{Carlin_2024}). This visual inspection campaign resulted in zero 3-vote candidates, two 2-vote candidates, 268 1-vote candidates, and 1,720 0-vote candidates, where ``X-vote'' means that a candidate was classified by X number of people as a potential dwarf galaxy. Only those candidates receiving two or more votes were considered our most probable dwarf galaxy detections.
Follow-up imaging for one of these two candidates was obtained using the Gemini Multi-Object Spectrograph (GMOS; \citealt{Hook2004}) on the Gemini South telescope, however these new data suggest that this candidate is not a real satellite of NGC~55. While no additional data were collected for the second candidate, its similarity to the first candidate suggests that it is likely either another background galaxy or galaxy cluster. See Appendix~\ref{appendix a} for more information. 

It should also be noted that while NGC~55-dw1 was detected by our resolved search method, it did not pass our visual inspection due to its extremely large size and LSB. We likely would have identified NGC~55-dw1 during visual inspection had we used a larger cutout size to generate candidate plots. % for each candidate. 
However, to optimize the inspection process for the more typical sizes of dwarf galaxies expected in our search, we adopted a smaller field of view (3~arcmin$^2$), which is better suited to these dwarfs, though too small to capture the full extent of NGC~55-dw1.

\section{Artificial Dwarf Galaxy Simulations} \label{sec:simulations}

\begin{figure*}[t]
    \centering
    \includegraphics[width = 0.8\textwidth]{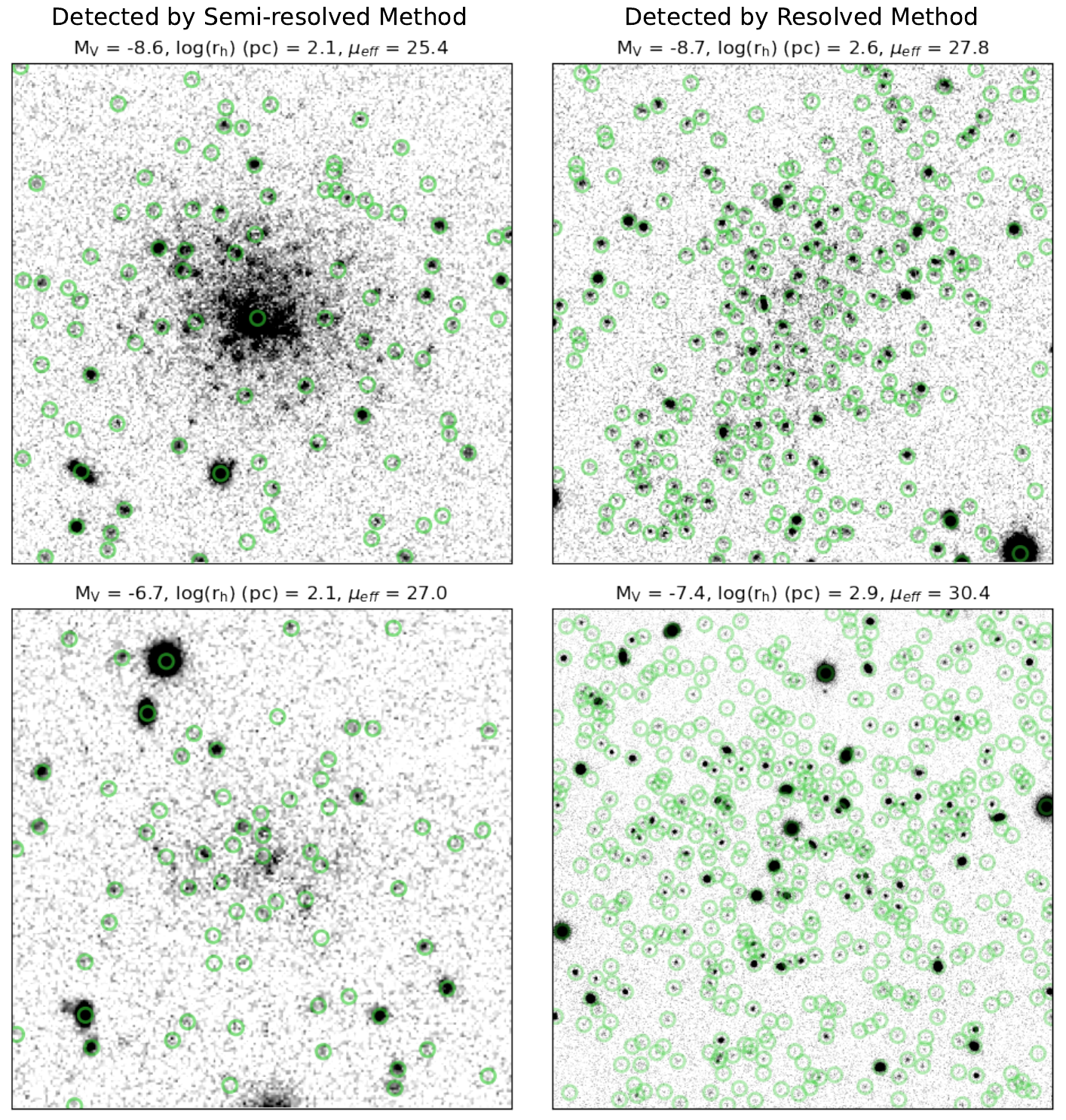}
    \caption{Four injected artificial dwarf galaxies of varying absolute magnitude, half-light radius, and effective surface brightness shown in $i$ band. Green circles indicate sources detected by \texttt{SourceExtractor}. Left column: Artificial dwarfs detected by our semiresolved method. Right column: Artificial dwarfs detected by our resolved method.} 
    \label{fig:artificial_dwarfs}
\end{figure*}

To properly interpret the results of our dwarf galaxy search, we need to characterize our detection depth and selection efficiency. To do this, we measure our dwarf galaxy detection sensitivity using artificial dwarf galaxy simulations. These simulations involve injecting artificial dwarfs into our data set and measuring the fraction of those that are recoverable. These injections are performed on the image level because dwarf galaxies at the distance of NGC~55 are not entirely resolved, and the unresolved portions of these galaxies can be more accurately modeled on the image level, as opposed to the catalog level (\citealt{Garling21}). To obtain a representative sample of the whole NGC~55 footprint, we inject dwarfs into 26 of the 184 DELVE-DEEP NGC 55 coadded tiles (or ``coadds'') that vary in exposure time and limiting magnitude and span the full range of depths covered by the DELVE-DEEP fields in Figure~\ref{fig:ngc55_depth}. 
 
To make sure our simulated galaxies are representative of real dwarfs, each simulated dwarf's stellar population is composed of old, metal-poor stars. Following \citet{Mutlupakdil21}, each stellar population is sampled from an isochrone modeling a galaxy at distance D$_{\mathrm{NGC55}}$ $\approx$ 2Mpc with an age of 10 Gyr and metallicity of [Fe/H] $=-2$, and generated according to a Salpeter IMF following $dN/dM \propto M^{-2.35}$ (\citealt{salpeter1955}). Isochrones are generated using the Dartmouth Stellar Evolution Database (\citealt{Dotter2008}). The stars within each dwarf's stellar population with $i \le 27$ mag are modeled as point sources convolved with the observed PSF and injected individually into the images according to an exponential profile. Rather than repeating this process for the faint, unresolved stars below our detection limits, all the stars with $i > 27$ mag are instead modeled and injected as a single exponential galaxy profile, such that each artificial dwarf is comprised of both a resolved and unresolved component. Both the point sources and exponential profiles are modeled using the galaxy image simulation package, \texttt{galsim} (\citealt{rowe2015galsim}). Our tests demonstrate that our dwarf-search pipeline detects these artificial dwarfs just as effectively as those in which all stars are injected as individual point sources.

A total of 2,287 artificial dwarfs, uniformly sampled from half-light radii $1.8 \le$ log($r_{h}$/pc) $\le 3.2$ and absolute magnitudes $-6.0 \ge M_V \ge -9.0$, are created and injected into the set of 26 coadds. To convert from $g$- and $i$-band magnitudes to $M_V$, we use the SDSS color conversions in \citet{Jordi_2006}, which give nearly the same results as the DES color conversions in \citet{DES_DR2}. Figure \ref{fig:artificial_dwarfs} shows four examples of artificial dwarf galaxies that are used in our analysis. We then treat the images with artificial dwarf galaxies in the same way as the unaltered images: each coadd is run through \texttt{SourceExtractor}, using the same parameters as in the DESDM processing of the real images.

\begin{figure*}[t]
    \centering
    \includegraphics[width = \textwidth]{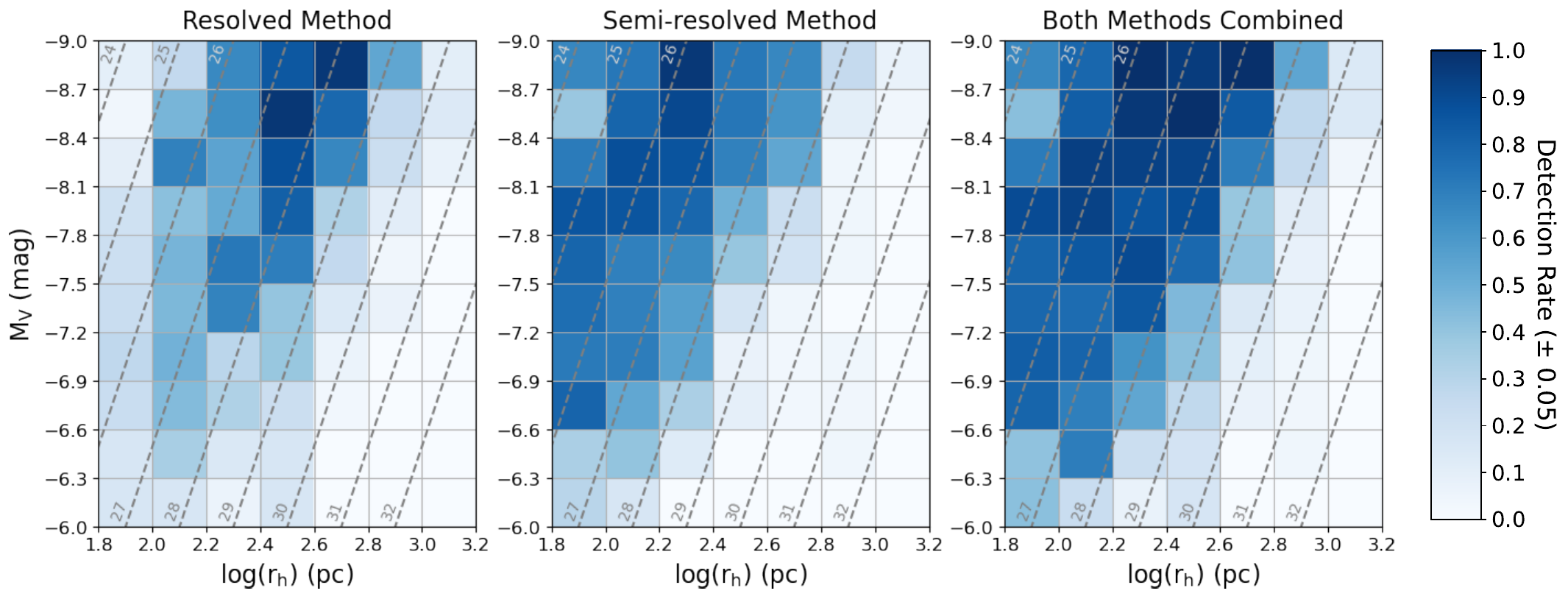}
    \caption{Dwarf galaxy detection sensitivity for each detection method individually (left and middle) and both methods combined (right). Detection rates have been corrected according to visual inspection results, where only artificial dwarfs that were classified as dwarfs by at least two people were counted as detections. Each bin contains, on average, 30 injected dwarfs with a standard deviation of 4. The corresponding average uncertainty in detection rate (calculated using binomial statistics) is 0.05. Lines of constant surface brightness are shown in gray (given in mag arcsec$^{-2}$). When convolved with the dwarf galaxy size-luminosity relation (\citealt{Brasseur_2011}), we are $\sim 80\%$ complete for $M_V \lesssim -7.8$ and $\sim 50\%$ complete for $M_V \lesssim -7.0$.}
    \label{fig:detection_sensitivity}
\end{figure*}

We run the resulting source catalogs through our dwarf-search pipeline (see Section~\ref{sec: dwarf_search}) to investigate our detection rate as a function of dwarf size ($r_h$) and luminosity ($M_V$). First, however, we account for dwarfs that, due to randomness in our injection process, are injected into regions of the NGC~55 halo that are already overdense and, as a result, are detected by our pipeline regardless of their parameters. To address this, our dwarf-search pipeline is run on both the altered images with injected dwarfs as well as the unaltered images. Any bins detected as overdensities in the unaltered images are removed from the dwarf search on the altered images, so that any artificial dwarfs located in these bins are removed from our sample and do not contribute to our detection sensitivity calculation.
%If an artificial dwarf is located in a bin that is already detected by our pipeline in the unaltered image, that dwarf is removed from our sample and does not contribute to our detection sensitivity calculation. 
After running our dwarf-search pipeline on each of the remaining artificial dwarfs, our team visually inspected each detection using Zooniverse. Just like in our real satellite search (see Section \ref{subsec:visual_inspection}), each dwarf was visually inspected by three team members, and only those identified as dwarfs by at least two of the three are considered detections. Both this Zooniverse inspection and that of the real satellite search were performed simultaneously to avoid any biases that could have resulted from inspecting the real candidates and artificial dwarfs separately.

Our resulting detection sensitivity (see Figure~\ref{fig:detection_sensitivity}) shows that the resolved detection method is more efficient for large, bright dwarf galaxies, while the semiresolved method is more sensitive to faint, compact ones (i.e., those with less resolved stars due to blending effects). With these two methods combined (through an outer join\footnote{Outer join means that the output tables of each detection method are combined such that all rows from both tables are included and duplicate rows are matched together. In \texttt{pandas.merge}, this corresponds to \texttt{how="outer"}.}; see right panel), we are sensitive to both regimes and can detect dwarf galaxies with $\gtrsim 80\%$ efficiency, on average, with effective surface brightness $\mu_{\text{eff}} \lesssim 28.5$ mag arcsec$^{-2}$ and $M_V \lesssim -6.6$. When these detection rates are convolved with the dwarf galaxy size-luminosity relation presented in \citet{Brasseur_2011}, we find that we are $\sim80\%$ complete to dwarf galaxies with $M_V \lesssim -7.8$. Detection rates drop for faint, diffuse dwarf galaxies (bottom right of each panel), as well as for very bright and compact dwarfs (top left corner of each panel), because these systems can no longer be identified as overdensities of individual stars or unresolved light components.
This highlights a known limitation in distinguishing between very compact, high surface brightness dwarfs and background galaxies in crowded fields.
It is also worth noting that, while our detection rates drop significantly for dwarfs with $\mu_{\text{eff}} \gtrsim 28.5$~mag arcsec$^{-2}$, many of these dwarfs (in particular, those with  $M_V \lesssim -7.5$) were detected by our resolved search method, but were not identified during the visual inspection due to their diffuse nature (similar to NGC~55-dw1; see Section~\ref{subsec:visual_inspection}). Additionally, while our sensitivity likely extends to dwarfs with log($r_{h}$/pc) $< 1.8$ ($r_{h} \lesssim 63$~pc), we chose not to explore this parameter space because there are currently very few known dwarf galaxies with $M_V \leq -6.0$ and $r_{h} \lesssim 63$~pc \citep{Mutlupakdil21,Pace_2022}. 

\section{Discussion}
\label{sec:discussion}

\begin{figure}[t]
    \centering
    \includegraphics[width = \columnwidth]{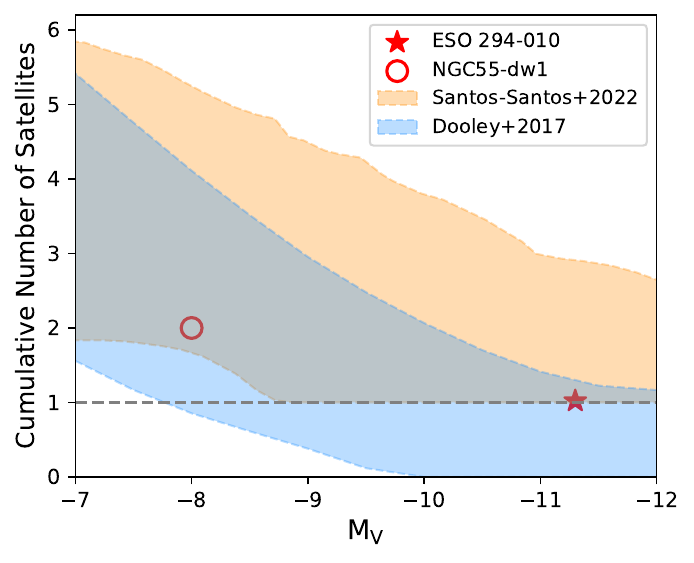}
    \caption{Satellite LF for NGC~55 down to $M_V = -7$ ($\sim50\%$ completeness limit of our dwarf search) given by the gray-dashed line. ESO 294-010 is shown with a red star and NGC~55-dw1 is shown with a red circle. While NGC~55-dw1 is a confirmed satellite of NGC 55, it is not included as part of our LF due to it being outside the size range of our sensitivity. The orange- and blue-shaded regions correspond to predicted satellite populations for an LMC-mass host, adapted from the 80\% scatter in \citet{Santos-Santos2022} and 1$\sigma$ scatter in \citet{Dooley2017b}, respectively. Both of these regions have been convolved with our observed sensitivity.}
    \label{fig:luminosity_function}
\end{figure}

We construct the satellite LF of NGC~55 that is roughly complete down to $M_V \sim -7$ and surface brightness $\mu_{\text{eff}} \sim 28.5$ mag arcsec$^{-2}$ (see Figure~\ref{fig:luminosity_function}). Our dwarf galaxy search (see Section~\ref{sec: dwarf_search}) yielded no new satellites beyond the two previously known systems, ESO 294-010 and NGC~55-dw1. Only the former of these systems is included as part of our LF because NGC~55-dw1 is an extremely diffuse dwarf galaxy ($\mu_{\text{eff}} \sim 32.3$ mag arcsec$^{-2}$; \citealt{McNanna2024}) and consequently falls below our completeness limits. It is because of this unusually diffuse nature that NGC~55-dw1 is not detected in our visual search (see Section \ref{subsec:visual_inspection}). We include ESO~294-010 in our LF since its luminosity and surface brightness fall within the assumed extent of our completeness. Although ESO~294-010 is located near the edge of NGC~55's virial radius, just beyond our search radius, its similar distance and radial velocity to NGC~55 (see Table~\ref{table:table1}) suggest that it is a likely satellite. As a result, our LF consists of one NGC~55 satellite to which we are complete.

We compare our LF to two existing $\Lambda$CDM-based predictions of the satellite populations of LMC-mass hosts (see \citealt{Carlin_2024} for similar comparisons with the NGC~2403 satellite system). The blue-shaded region in Figure \ref{fig:luminosity_function} corresponds to the 1$\sigma$ scatter in predicted satellites of an isolated, LMC-mass host, as presented in \citet{Dooley2017b}. These predictions are based on the \textit{Caterpillar} dark matter halo simulations (\citealt{Griffen_2016}) and stellar mass-halo mass relation of \citet{Garrison-Kimmel_2017}. The orange-shaded region represents the 10-90 percentile scatter in predicted satellites around each of the DELVE-DEEP MC analogs, from \citet{Santos-Santos2022} (specifically, this region corresponds to the “cut-off” model from \citealt{Santos-Santos2022}). It should be noted that this model starts counting the cumulative number of satellites at~1 rather than~0, thereby excluding the possibility that such hosts have no satellites. These predictions are based on the Local Group cosmological simulation, APOSTLE \citep{Fattahi_2016}.
In both the \citet{Dooley2017b} and \citet{Santos-Santos2022} simulations, only subhalos within the virial radius of each primary halo were considered as potential satellites, making the corresponding predictions directly comparable to our LF for NGC 55, since our search was also restricted to the host's virial radius.

Both the blue and orange regions have been convolved with our observed sensitivity (Figure~\ref{fig:detection_sensitivity}) and thus represent the predicted number of observed satellites rather than the number of satellites predicted by simulations. To perform these convolutions, it was necessary to determine our sensitivity as a function of only $M_V$, as opposed to both $M_V$ and $r_h$, which was achieved using the dwarf galaxy size-luminosity relation presented in \citet{Brasseur_2011}. With our sensitivity taken into account, our LF aligns well with both $\Lambda$CDM predictions. On the bright end it agrees closely with each of the predicted regions, while on the faint end it is slightly underpopulated.

Our LF is also in agreement with two recent LFs for the LMC analog, NGC~2403 (\citealt{Carlin_2024}), and SMC analog, NGC~3109 (\citealt{Doliva-Dolinsky_2025}), each of which contain two satellites with $-7~\geq~M_V~\geq~-13$ around their respective hosts, and are consistent with $\Lambda$CDM predictions. Additionally, our LF is consistent with the results of two recent dwarf galaxy surveys targeting MC analogs at distances of $4-10$ Mpc: ID-MAGE \citep{Congreve_2025} and ELVES-Dwarf \citep{Li_2025}. ID-MAGE reports an average of $4.0\pm1.4$ satellite candidates per LMC-mass host (with a lower bound of $1.4\pm0.6$ satellites), and ELVES-Dwarf finds $0-2$ satellites around each of their MC-mass hosts. Both of these bounds agree well with NGC~55 having two satellites. It is worth noting, however, that NGC~55 is significantly closer than the hosts targeted by ID-MAGE and ELVES-Dwarf. While these complementary surveys rely exclusively on integrated light detection—allowing them to probe a larger number of hosts—, they are limited in their ability to detect the faintest satellites. As a result, both surveys are complete only down to $M_V \sim -9$, roughly two magnitudes brighter than our completeness limit. The relative proximity of NGC~55 enables (semi)resolved star searches in our study, allowing us to reach significantly fainter systems that would otherwise remain undetectable.

%ID-MAGE reports a lower estimate of $1.4\pm0.6$ satellite candidates per LMC-mass host \citep{Congreve_2025}, and ELVES-Dwarf finds $0-2$ satellites around each of their MC-mass hosts \citep{Li_2025}. Both of these bounds agree well with NGC~55 having two satellites. It is worth noting, however, that NGC~55 is much closer than the hosts studied in these surveys. While resolved star searches are still possible for NGC~55, these surveys must rely solely on integrated light searches, which can probe more hosts but cannot reach the faintest systems detectable in our search.
%As a result, both ID-MAGE and ELVES-Dwarf are complete down to $M_V \sim -9$, which differs from our completeness by $\sim 2$ mag.}
% This means their satellite population estimates exclude the faintest regime explored in our dwarf search.

\begin{figure}[t]
    \centering
    \includegraphics[width = \columnwidth]{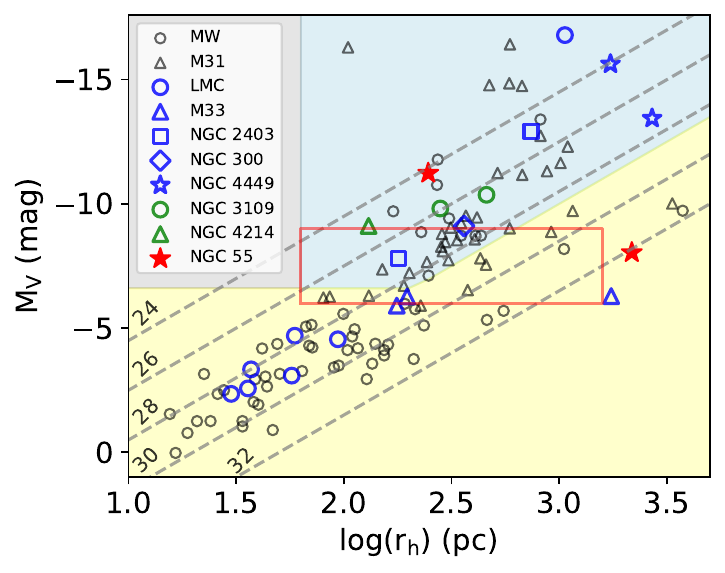}
    \caption{Absolute magnitude vs. half-light radius for known dwarf satellites. Open gray symbols indicate MW and M31 satellites from the compilation of \citealt{Pace_2024}. Open blue symbols indicate satellites of the LMC (\citealt{Patel_2020}) or LMC analogs: M33 (\citealt{Chapman2013}; \citealt{Mart_nez_Delgado_2021}; \citealt{collins2024}; \citealt{ogami2024}), NGC~2403 (\citealt{Carlin_2024}), NGC~300 (\citealt{sand_2024}), NGC~4449 (\citealt{Rich_2012}; \citealt{Martinez_Delgado_2012}). Open green symbols indicate satellites of SMC analogs: NGC~3109 (\citealt{Sharina2008}; \citealt{Sand15b}; \citealt{Doliva-Dolinsky_2025}), and NGC~4214 (\citealt{carlin21}). Closed red stars indicate NGC~55 satellites. The red rectangle outlines the parameter space that was tested in artificial dwarf simulations (Section \ref{sec:simulations}). The light blue-shaded region indicates where our dwarf search is $\gtrsim 80\%$ complete, and the yellow-shaded region indicates where our search is $\lesssim 80\%$ complete. Completenesses in these regions outside of the red rectangle are extrapolated. The gray-shaded region indicates where we cannot extrapolate our completeness. Gray-dashed lines indicate lines of constant surface brightness.}
    \label{fig:known_dwarfs}
\end{figure}

In Figure~\ref{fig:known_dwarfs}, we place the two dwarf satellites of NGC~55 in context with other MC analogs in the size-luminosity plane. NGC~55's brighter satellite companion, ESO~294-010, is consistent with other known dwarfs in terms of size and luminosity ($M_V \sim -11.3$; $r_h \sim 248$~pc). In addition to an old, metal-poor stellar population, ESO~294-010 contains a small population of young, main-sequence stars as well as \hi~gas (\citealt{Karachentsev_2002}; \citealt{Da_costa_2010}). NGC~55-dw1, on the other hand, is similar in magnitude to a number of other known dwarfs, but is unusually large for its luminosity ($M_V \sim -8.0$; $r_h \sim 2.2$~kpc), which is possibly a result of tidal interactions with NGC~55 \citep{McNanna2024}. Additionally, the lack of deeper imaging around this dwarf makes it difficult to conclude whether or not it contains any young stellar populations. That said, NGC~55 appears to have a distorted \hi~disk with some gas extending in the direction of NGC~55-dw1, which could potentially be a sign of tidal interaction (\citealt{Westmeier_2013}). Furthermore, the unusually diffuse nature of NGC~55-dw1 raises the question of whether it is truly an intact system or part of a larger tidal debris structure. While similarly diffuse dwarfs, such as Antlia~II and Crater~II, are generally regarded as gravitationally bound systems (\citealt{Torrealba_2016, Torreabla_2019}; \citealt{Ji_2021}), the lack of dynamical observations for NGC~55-dw1 leaves its bound nature uncertain.

\section{Conclusion} \label{sec:conclusions}

In this work, we perform a systematic satellite search around a nearby LMC-mass galaxy, NGC~55 (2.1 Mpc). Unlike previous satellite searches around low-mass hosts, our approach combines both resolved and semiresolved detection methods. As a result, our satellite search yields two potential dwarf galaxy candidates, although follow-up GMOS imaging for one of these candidates indicates that it is not a satellite of NGC~55, and the second candidate's similarity to the first suggests that it is not a satellite either (see Appendix~\ref{appendix a}). We present a satellite LF for NGC~55 complete down to $M_V \sim -7$ that consists of one dwarf galaxy, ESO~294-010, with $M_V \sim -11.3$ (see Figure~\ref{fig:luminosity_function}). We do not include NGC~55's second known satellite, NGC~55-dw1, in our LF because its unusually diffuse nature places it outside our completeness limits. Nevertheless, our LF is in good agreement with the $\Lambda$CDM predictions presented in \citet{Dooley2017b} and \citet{Santos-Santos2022} and with the results of recent dwarf galaxy searches around other MC analogs (\citealt{Carlin_2024}, \citealt{Congreve_2025}, \citealt{Doliva-Dolinsky_2025}, \citealt{Li_2025}). Furthermore, when placed in context with other satellites of such MC analogs, ESO~294-010 demonstrates a size and luminosity consistent with those of similar systems, while NGC~55-dw1 is notably larger in comparison to other dwarfs with similar luminosities. This unusually extended structure could suggest that NGC~55-dw1 has undergone tidal disruption by NGC~55 or that it is not a completely intact satellite.

In determining the completeness of our satellite search, our artificial dwarf galaxy simulations show that the combination of resolved and semiresolved search methods results in the detection of a broad range of dwarf galaxy properties, since we are sensitive to dwarfs with $\mu_{\text{eff}} \lesssim 28.5$ mag arcsec$^{-2}$ and $M_V \lesssim -6.6$ and $\sim 80\%$ complete to $M_V \lesssim -7.8$ (see Figure~\ref{fig:detection_sensitivity}). This approach could therefore enhance the completeness and robustness of future satellite searches targeting a similar regime of dwarf galaxies. Machine learning techniques could be particularly adept at integrating both resolved and unresolved information to enhance the efficiency of such searches (e.g., \citealt{Muller_2021}, \citealt{Tanoglidis2021}, \citealt{Jones_2023}).

As one of the only comprehensive satellite studies around a low-mass host, our results emphasize the need to rigorously characterize additional MC-analog satellite systems. With recent and upcoming papers (\citealt{Carlin_2024}; \citealt{Doliva-Dolinsky_2025}) beginning to increase our sample of explored MC-mass hosts, the DELVE-DEEP survey will conduct satellite searches around its three remaining targets, further improving the characterization of satellites around low-mass hosts in the Local Volume. DELVE-DEEP will also perform systemic searches for substructures around the remaining hosts, as has recently been done for NGC~300 \citep{Fielder_2025}. To complement this, the MADCASH survey will execute satellite and substructure searches around its five remaining MC-mass targets \citep{Carlin_2024}. Within the next year, the Vera C. Rubin Observatory's Legacy Survey of Space and Time (LSST; \citealt{Ivezic_2019}) will begin operations. LSST will likely be able to identify hundreds of new dwarf galaxies in the Southern sky, including both satellites of larger galaxies and isolated field dwarfs, allowing for a better evaluation of the $\Lambda$CDM model on small scales \citep{Mutlupakdil21}. Dwarf searches performed in the LSST era could also benefit from combining resolved and semiresolved search methods, as demonstrated in this work. This approach may enable the detection of more, and even fainter, dwarf galaxies than current predictions suggest.

\section{Acknowledgements}
\label{sec:acknowledgements}

We thank the anonymous reviewer for their careful reading of our manuscript and their insightful comments and suggestions. J.M. acknowledges support from the University of Chicago Quad Undergraduate Research Scholars Program. W.C. gratefully acknowledges support from a Gruber Science Fellowship at Yale University. This material is based upon work supported by the National Science Foundation Graduate Research Fellowship Program under Grant No. DGE2139841. Any opinions, findings, and conclusions or recommendations expressed in this material are those of the author(s) and do not necessarily reflect the views of the National Science Foundation.

This material is based upon work supported by the National Science Foundation under Grant No. AST-2108168, AST-2108169, AST-2307126, and AST-2407526. This research award is partially funded by a generous gift of Charles Simonyi to the NSF Division of Astronomical Sciences. The award is made in recognition of significant contributions to Rubin Observatory’s Legacy Survey of Space and Time.

The DELVE Survey gratefully acknowledges support from Fermilab LDRD (L2019.011), the NASA Fermi Guest Investigator Program Cycle 9 (No. 91201), and the National Science Foundation (AST-2108168, AST-2307126). This work was supported in part by the U.S. Department of Energy, Office of Science, Office of Workforce Development for Teachers and Scientists (WDTS) under the Science Undergraduate Laboratory Internships Program (SULI).

This publication uses data generated via the Zooniverse.org platform, development of which is funded by generous support, including a Global Impact Award from Google, and by a grant from the Alfred P. Sloan Foundation.

This project used data obtained with the Dark Energy Camera (DECam), which was constructed by the Dark Energy Survey (DES) collaboration. Funding for the DES Projects has been provided by the DOE and NSF (USA), MISE (Spain), STFC (UK), HEFCE (UK), NCSA (UIUC), KICP (U. Chicago), CCAPP (Ohio State), MIFPA (Texas A\&M), CNPQ, FAPERJ, FINEP (Brazil), MINECO (Spain), DFG (Germany) and the Collaborating Institutions in the Dark Energy Survey, which are Argonne Lab, UC Santa Cruz, University of Cambridge, CIEMAT-Madrid, University of Chicago, University College London, DES-Brazil Consortium, University of Edinburgh, ETH Zürich, Fermilab, University of Illinois, ICE (IEEC-CSIC), IFAE Barcelona, Lawrence Berkeley Lab, LMU München and the associated Excellence Cluster Universe, University of Michigan, NOIRLab, University of Nottingham, Ohio State University, OzDES Membership Consortium, University of Pennsylvania, University of Portsmouth, SLAC National Lab, Stanford University, University of Sussex, and Texas A\&M University.

Based on observations at Cerro Tololo Inter-American Observatory at NSF’s NOIRLab (NOIRLab Prop. ID 2019A-0305; PI: A. Drlica-Wagner), which is managed by the Association of Universities for Research in Astronomy (AURA) under a cooperative agreement with the National Science Foundation.

Based on observations obtained at the international Gemini Observatory (GS-2024B-FT-204; PI: J. Medoff), a program of NSF NOIRLab, which is managed by the Association of Universities for Research in Astronomy (AURA) under a cooperative agreement with the U.S. National Science Foundation on behalf of the Gemini Observatory partnership: the U.S. National Science Foundation (United States), National Research Council (Canada), Agencia Nacional de Investigaci\'{o}n y Desarrollo (Chile), Ministerio de Ciencia, Tecnolog\'{i}a e Innovaci\'{o}n (Argentina), Minist\'{e}rio da Ci\^{e}ncia, Tecnologia, Inova\c{c}\~{o}es e Comunica\c{c}\~{o}es (Brazil), and Korea Astronomy and Space Science Institute (Republic of Korea).

\vspace{5mm}
\facilities{Blanco}

\software{Astropy \citep{astropy13,astropy18}}

\bibliographystyle{aasjournal}
\bibliography{reference}

%\clearpage

\appendix
\label{appendix}

\section{Stellar Completeness}
\label{appendix b}
To measure our stellar completeness, we inject artificial stars into our data set and measure the fraction of those that are recoverable. Like in Section \ref{sec:simulations}, stars are injected on the image level and into the same set of 26 NGC~55 coadds.

Artificial stars are modeled as point sources convolved with the observed PSF using the galaxy image simulation package, \texttt{galsim} (\citealt{rowe2015galsim}). Square grids of 5,000-10,000 stars are injected into each of the 26 coadds with random g-band magnitudes between $21 \le g \le 27$ mag. To avoid significantly increasing the average background magnitude, the exact number of stars injected into each coadd was $\sim$10\% of the number of real sources in the coadd. The colors of the artificial stars (and as a result, their $r$ and $i$ magnitudes) in each coadd are randomly sampled from the color distribution of real sources in that coadd ($-0.50 \lesssim g-r \lesssim 1.25$ and $-0.50 \lesssim g-i \lesssim 2.00$). This process is repeated for each coadd 10 times, resulting in a total of 260 coadds containing injected stars. Each coadd is then run through \texttt{SourceExtractor} for recovery, using the same parameters as in the DESDM processing of the real images. 

\begin{figure}[H]
    \centering
    \includegraphics[width = 0.7\columnwidth]{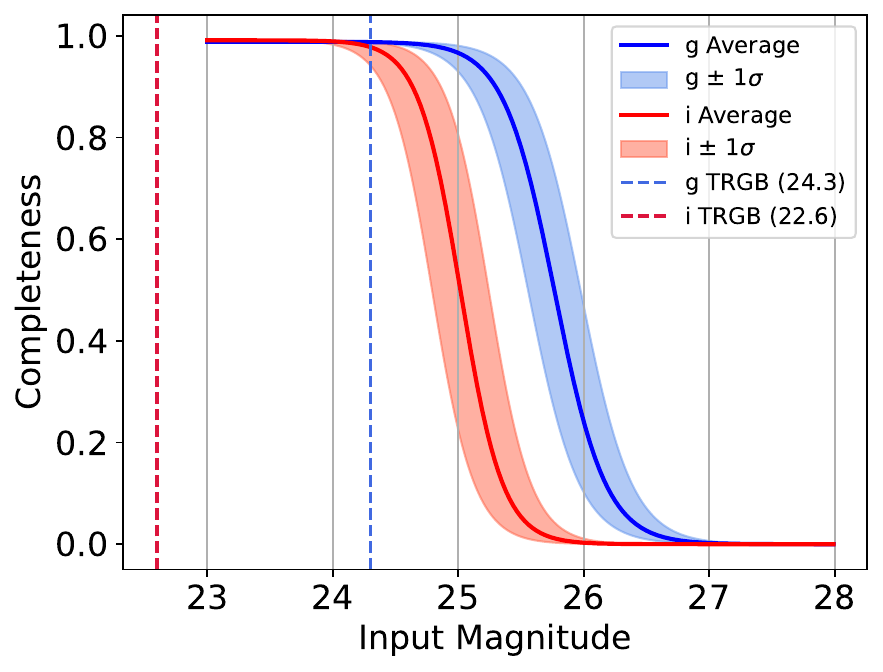}
    \caption{Average completeness in $g$ and $i$ bands across all 26 coadds used for the artificial star injections. Solid lines indicate mean completeness and shaded regions indicate $\pm1\sigma$. Average 50\% completeness is $25.8 \pm 0.2$ mag in $g$ and $25.0 \pm 0.2$ mag in $i$. TRGB apparent magnitudes in $g$ and $i$ given by blue- and red-dashed lines, respectively ($m_g^{\mathrm{TRGB}} \sim 24.3$; $m_i^{\mathrm{TRGB}} \sim 22.6$).}
    \label{fig:completeness}
\end{figure}

After running this pipeline on the coadds containing artificial sources, the completeness of each coadd is determined by calculating the fraction of injected stars that are recovered as a function of input magnitude. Since we are not applying star-galaxy separation, stars are considered recovered as long as they are detected by \texttt{SourceExtractor} regardless of whether they are classified as point sources. Our measured completenesses are then fitted according to the completeness model described in Eq. 7 of \citealt{Martin2016}, which takes the form of a decreasing logistic function. The average completenesses across the 26 tested coadds are shown in Figure \ref{fig:completeness}. 

Our tests show that we are, on average, 90\% complete to $g = 25.3 \pm 0.2$ mag and $i = 24.6 \pm 0.2$ mag, which are $\sim1.0$ and $\sim2.0$ mag fainter than the corresponding TRGB magnitudes, $m_g^{\mathrm{TRGB}} \sim 24.3$ and $m_i^{\mathrm{TRGB}} \sim 22.6$ (see Figure \ref{fig:completeness}). These TRGB magnitudes are estimated using an old, metal-poor isochrone with [Fe/H] $=-2$ and age = 10 Gyr at the distance of NGC~55. Isochrones are generated using the Dartmouth Stellar Evolution Database (\citealt{Dotter2008}). 

\section{Two Satellite Candidates}
\label{appendix a}

\begin{figure*}[b]
    \centering
    \includegraphics[width = \textwidth]{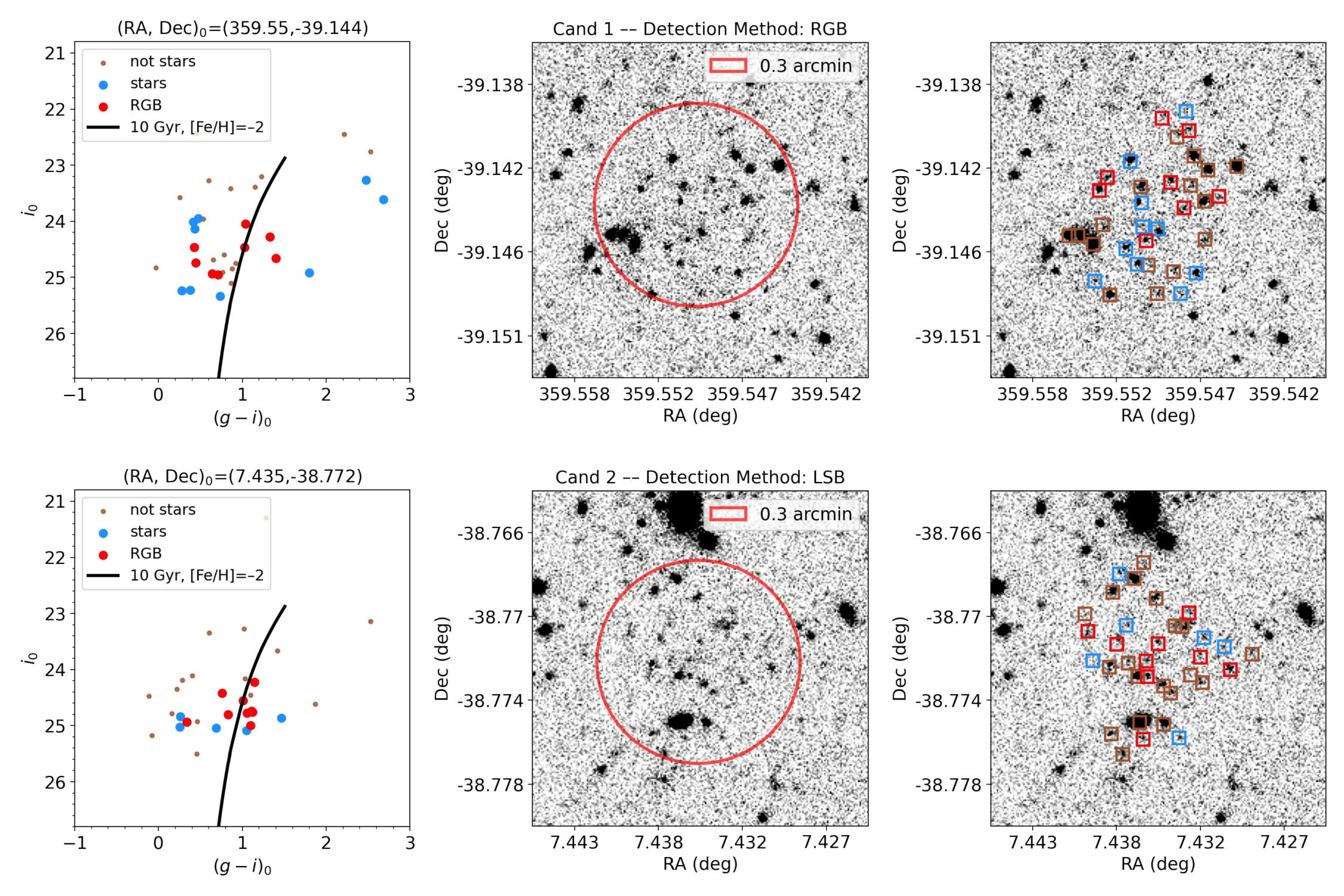}
    \caption{The DELVE-DEEP color-magnitude diagram of our 2-vote candidates (top: Candidate~1, bottom: Candidate~2) and their $i$-band images. RGB stars are shown in red, stars outside of our RGB selection region are shown in blue, and all other sources are shown in brown. The black line represents an old, metal-poor isochrone (age $= 10$~Gyr, [Fe/H]$=-2$, \citealt{Dotter2008}) at the distance of NGC~55. The left panel includes only the sources within a radius of 0.3~arcmin surrounding the candidate, indicated by the red circle in the middle panel. The corresponding detection method is given in the title. The right panel highlights each source from the left panel on the $i$-band image.    }
    \label{fig:both_cand_plots}
\end{figure*}

Figure~\ref{fig:both_cand_plots} shows DELVE-DEEP images and color-magnitude diagrams of each of the 2-vote candidates that resulted from our satellite search. Note that the first of these candidates is detected by our resolved search method, whereas the second is detected by our semiresolved search method, despite both exhibiting overdensities of resolved stars, which is why we include a color-magnitude diagram for each. Brown points indicate sources that are not classified as stars, blue points represent sources identified as stars through star-galaxy separation, and red points highlight stars within our RGB selection region. Only the sources within a 0.3~arcmin radius from the candidate's center, marked by a red circle in the middle panel, are shown in the left panel. These sources are highlighted with corresponding colors on the $i$-band image in the right panel. 

We obtained the follow-up imaging for the first of these two candidates with the Gemini Multi-Object Spectrograph (GMOS; \citealt{Hook2004}) on the Gemini South telescope (GS-2024B-FT-204; PI: J. Medoff). The GMOS images have a $\sim$5.5\arcmin$\times$5.5\arcmin field of view and 0.16\arcsec pixel$^{-1}$ scale after binning. Both $g$ and $i$-band imaging was taken with strict image quality constraints on 2024 December 25 (UT). We collected 9$\times$300~s $g$-band exposures and 4$\times$300~s $i$-band exposures, with small dithers between each exposure. The goal of these follow-up observations was to increase the number of putative member stars in our candidate system in order to confirm whether they represented bona fide associations of stars at the distance of NGC~55.

Initial data reduction was conducted using DRAGONS \citep{Labrie2023}, the pipeline maintained by Gemini Observatory. DRAGONS performs bias subtraction, flat-field correction, and bad pixel masking on the images. Cosmic rays were removed using the sigma-clipping method within the DRAGONS pipeline. Stacked images were created using the weighted average of the individual exposures. An astrometric correction was applied by using SCAMP \citep{Bertin2010scamp}. The final $g$- and $i$-band stacked images had PSF FWHM values around 0.7\arcsec.

\begin{figure*}[h]
    \centering
    \includegraphics[width = \textwidth]{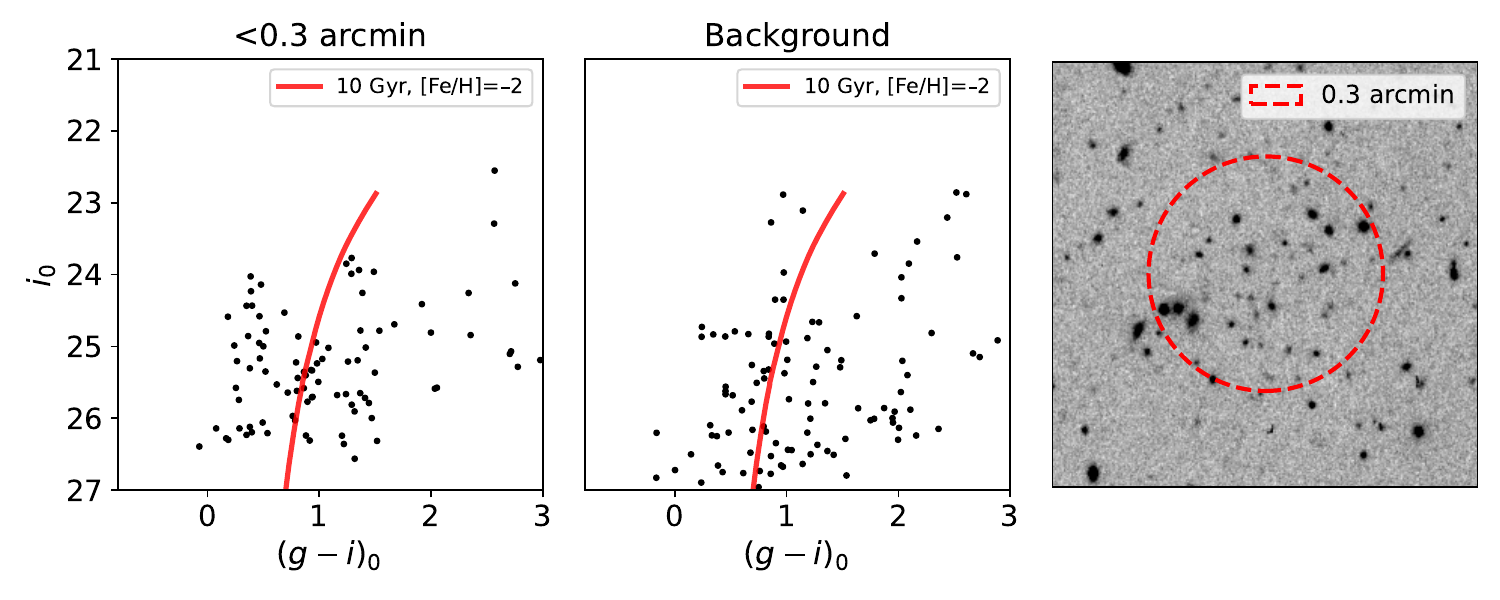}
    \caption{Follow-up Gemini data for Candidate 1. Left: Deep color-magnitude diagram of the region within a 0.3~arcmin radius of the candidate. Isochrone with age $= 10$ Gyr and metallicity [Fe/H] $=-2$ at the distance of NGC~55 is shown in red (\citealt{Dotter2008}). Middle: Same as left but for a background region of equal area. Right: An $i$-band image with a dashed red circle of a radius 0.3~arcmin surrounding the candidate.}
    \label{fig:gemini}
\end{figure*}

We performed PSF fitting photometry on the stacked GMOS images, using DAOPHOT and ALLFRAME \citep{Stetson87,Stetson94}, following the general procedure described in \citet{MutluPakdil2018}. The photometry was calibrated to point sources in our DELVE-DEEP catalog, including a color term, and was corrected for Galactic extinction \citep{Schlafly11} on a star-by-star basis. In Figure~\ref{fig:gemini}, we show the deep CMD of candidate~1 within a radius of 0.3\arcmin, along with a representative background field of the same area. The overplotted red line represents an old, metal-poor isochrone (age $= 10$ Gyr; [Fe/H] $=-2$; \citealt{Dotter2008}) at the distance of NGC~55. The absence of stars near the isochrone and the overwhelming presence of background sources suggest that the candidate is not a real system at the distance of NGC 55.

No follow-up data were obtained for the second candidate (see Figure~\ref{fig:both_cand_plots}-bottom panels), but its similarity in appearance and the CMD to the first candidate suggests that it is unlikely to be a real system at the distance of NGC~55. In addition, both candidates are visible in the Wide-field Infrared Survey Explorer (WISE) data, specifically the unWISE W1/W2 NEO7 images. This may indicate that they are, instead, background, high-redshift galaxy clusters.

\end{document}